Caroline O. Buckee and Kenth Engø-Monsen

# Mobile phone data for public health: towards data-sharing solutions that protect individual privacy and national security

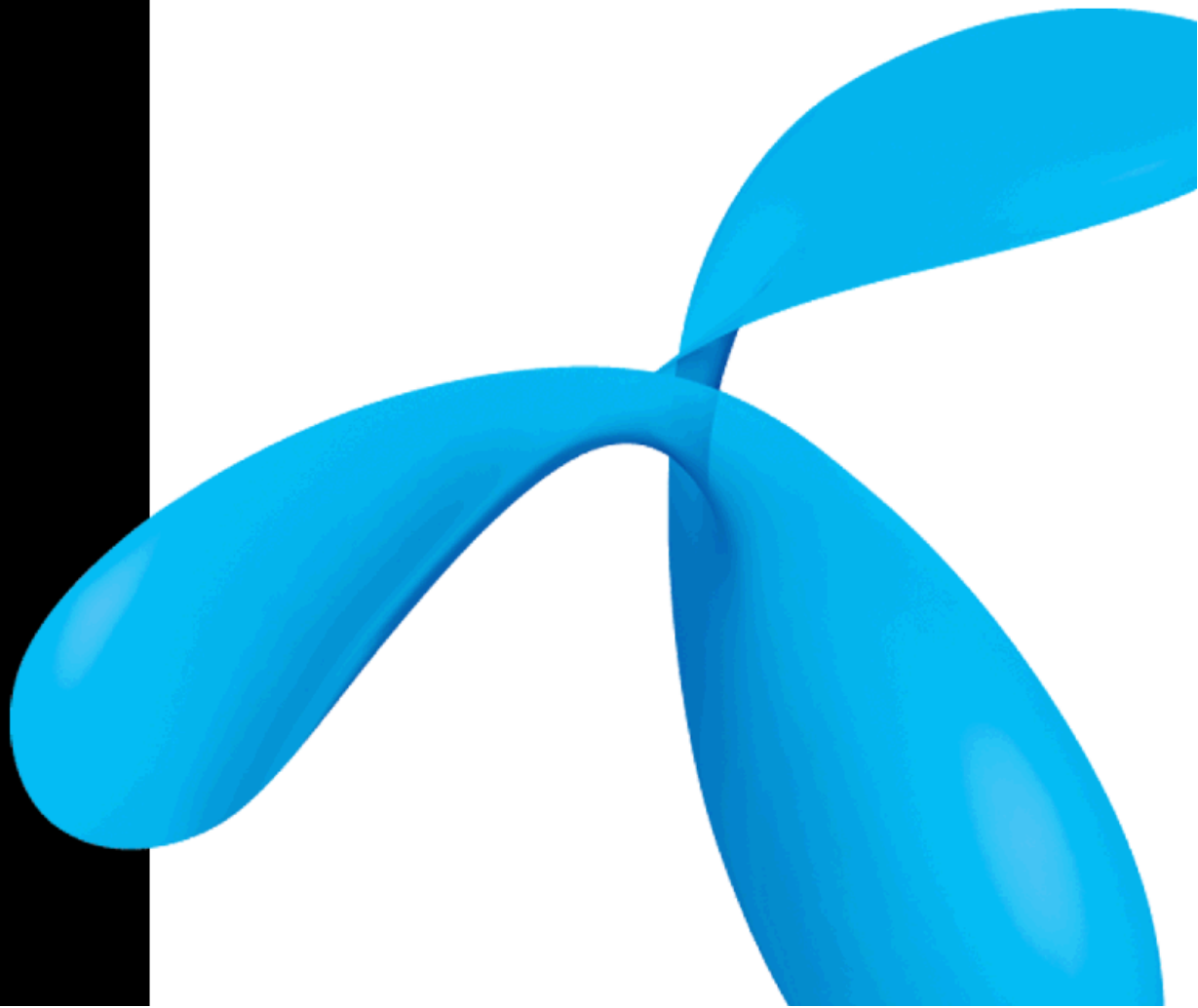

| **Telenor report** | **2 / 2016** |
| --- | --- |
| **Title** | **Mobile phone data for public health: towards data-sharing solutions that protect individual privacy and national security** |

| **Author(s)** | Caroline O. Buckee and Kenth Engø-Monsen |
| --- | --- |
| | Affiliation Caroline O. Buckee: |
| | Department of Epidemiology, Harvard T.H. Chan School of Public Health, USA |
| | Center for Communicable Disease Dynamics, Harvard T.H. Chan School of Public Health, USA |
| | Affiliation Kenth Engø-Monsen: |
| | Telenor Research, Telenor Group, Norway |

## Abstract

We outline the constraints faced by operators when deciding to share de-identified data with researchers or policy makers. We describe a conservative approach that we have taken to harness the value of CDRs for infectious disease epidemiology while ensuring that identification of individuals is impossible. We believe this approach serves as a useful and highly conservative model for productive partnerships between mobile operators, researchers, and public health practitioners.

# Contents

# 1 Background

Quantifying populations and understanding how they interact on different spatial and temporal scales is essential for many development and public health policy challenges. The rapid adoption of mobile technologies around the world, even among the poorest and least accessible communities, has led to widespread excitement about the use of mobile phone data to observe and engage with hard-to-reach populations[1-5]. In particular, a rapidly growing field of research is emerging around the analysis of call detail records (CDRs): the "digital traces" that mobile telecommunication operators store for billing purposes. CDRs provide a record of the cell tower IDs, and thus approximate locations, of subscribers whenever their mobile phone is used. CDRs – if appropriately de-identified to protect individual privacy – can therefore be used to model the location and movements of millions of individuals over time, and have been used in behavioural, economic, and epidemiological studies, as well as in the wake of natural disasters[6-17].

Before these data sets, census data and small travel surveys represented the most reliable sources of information about human population distributions and mobility patterns in low-income settings. Although rich in individual-level information, these often fail to cover relevant timescales for public health challenges like epidemics, are not updated frequently, and may not provide sufficient scale. On the other hand, although de-identified CDRs lack individual-level metadata, they provide scalable, longitudinal measures of populations in near real-time. Despite the clear potential of their data for social good, however, most mobile operators are extremely reluctant to even consider providing access to researchers and policy makers. Important (albeit theoretical) privacy concerns and risks of re-identification have been raised, and understandably these make both mobile operators and national regulators very wary.

We outline the constraints faced by operators when deciding to share de-identified data with researchers or policy makers. We describe a conservative approach that we have taken to harness the value of CDRs for infectious disease epidemiology while ensuring that identification of individuals is impossible. We believe this approach serves as a useful and secure model for productive partnerships between mobile operators, researchers, and public health practitioners.

# 2 Matching data sharing agreements to analysis requirements

Different public health applications require different levels of spatial and temporal data resolution. Many critical policy questions require the mapping of population distributions and regional movement patterns, and this level of resolution does not need individual-level detail. Here, spatially aggregated data reflecting administrative units on daily or weekly timescales may be ideal. For example, to assess population displacement and routes of migration following natural disasters[17], to understand the general mobility patterns of different communities, or to measure the flow of populations between regions that spread disease[11,12,16], even relatively coarse grained information about population numbers within, and the rate of flow between, different administrative units is immensely helpful. On the other hand, to understand how individual commuting patterns within a city might spread dengue fever or expose different communities to pollution, particular trajectories are important[18]. Since operators in different locations are likely to be subject to different regulations, the scope of CDR-based analyses will also vary. Data sharing agreements must therefore be carefully tailored to the specific application and region.

# 3 Stakeholder perspectives on data sharing

By accepting the Terms and Conditions of a mobile phone contract, individual subscribers agree to the analysis of their personal data for the purposes of service delivery and network security issues, but seldom for research purposes. Given the possibility of re-identification of individuals from CDRs, resulting from the predictability of peoples' routines, protecting personal information from researchers or governments is crucial, and should form the basis of data-sharing agreements, as it does for national regulators that award licenses - known as the telecom concession – that cover all aspects of operators' data processing. Pronounced country-specific differences in the concession exist due to national legislation, which is designed to ensure that privacy and state security measures are met by the mobile operators when they collect and process personal information. In some countries, for example, national security legislation constrains the export of personal data from the mobile operator across the border to another country, and all processing and handling of the detailed personal data must be done in-country.

From the mobile operator's perspective, CDRs are primarily analyzed in order to bill subscribers, deliver services, and maintain and secure the network. Even if regulators approve the use of CDRs for research purposes, as commercial businesses mobile operators may see no value in making an investment in de-identifying or aggregating CDRs, which is often a non-trivial exercise, given the sheer size of data sets. Typically, a mobile operator can collect billions of CDRs each day, so longitudinal studies can easily require 10's to 100's of terabytes. Simply handling such amounts of data raises computational challenges that require significant investment of personnel by operators. Given the perceived risks of outside parties analyzing sensitive data and the investment required of mobile operators to undertake data-sharing, it is perhaps unsurprising that in the absence of standard protocols, gaining access to CDRs for public health purposes - even in response to emergencies such as the Ebola virus disease outbreak - is extremely difficult for researchers[16].

We propose that there are also advantages for mobile operators to collaborate with researchers and public health agencies, however. First, there are benefits for the mobile operator in the potential 'analytical spill-over' effects in explorative research work done in collaboration with academic partners that can aid in the everyday running of business. For example, insights generated in research projects may feed directly into the churn and retention activities, or into marketing campaigns. Second, there are reputational and indirect business effects through corporate social responsibility activities when operators collaborate with researchers on solving humanitarian or social problems. Third, in the long-term there is also value in building the capabilities of governments and international organizations to address socioeconomic challenges, by enhancing customers' trust, increasing transparency, and strengthening societies by building and supporting sustainable infrastructure solutions. We believe there are low-cost ways for mobile operators to undertake CDR-based projects, particularly if protocols are in place that ensures a high standard for privacy protection and funding for necessary privatization steps can be paid for from grants by researchers or development agencies. Given the potential of these data sets for social good, it is important to develop models of data sharing that achieve these goals.

# 4 A conservative model of data sharing

While other types of "Big Data" also have important privacy implications that hinder simple sharing arrangements, not least medical record data, genomic data, and certain types of social media data, for the telecom industry, standardized frameworks do not currently exist. As a result, agreements are not standardized, but rather highly variable and ad-hoc. Guidelines are often opaque and highly political, so negotiating data access can be a lengthy, highly variable process requiring various government agencies' approval and regulator support. We have recently conducted a collaboration between Telenor Research, the mobile operator Telenor Pakistan, and Harvard T.H. Chan School of Public Health, that we believe can serve as a highly conservative model for the analysis of CDRs for public health. Telenor Research and Telenor Pakistan are part of the Telenor Group, a corporate group of mobile operators originating in Norway, and with a current customer footprint in Scandinavia, Eastern Europe and South-East Asia. The research question was focused on the role of human mobility in the recent emergence and spread of dengue virus[11].

To overcome issues relating to privacy and security we worked together to formulate a protocol based on the following principles. Note that these are more conservative than would be ideal for research questions that require individual-level detail, but they highlight the fact that it is possible to avoid privacy concerns if the specific application is matched appropriately to the data sharing agreement:

**Accessing data and security of CDRs:** Raw CDRs – which consist of rows of SIM card/phone numbers, time and date stamp, cell tower ID – remained within the operator's data warehouse and were processed only by employees of the operator's in-country business unit. This ensured that researchers never had direct contact with sensitive data and CDRs did not leave the country, in accordance with national regulations.

**Processing CDRs:** SIM card/phone numbers were de-identified using hashed ID values for SIM IDs by Telenor Pakistan employees prior to aggregation by Telenor Research employees. Derived, and processed and aggregated data sets were stored at Telenor Research's servers to ensure that the operator maintained control over access and aggregation, even of anonymized data, for future projects.

**Suitability of aggregation for research question:** Researchers worked with Telenor Research in-country to specify the aggregation requirements for the study, which were specific to the research question. Aggregation in this case refers to measuring the total number of subscribers located at each tower per day, and aggregate number of transitions between towers occurring per day. This data can be further spatially or temporally aggregated (e.g. total transitions between towers located in City A and towers located in City B per week).

**Analysis of aggregated matrices:** Telenor Research then provided access to spatially and temporally aggregated matrices to Harvard researchers for subsequent analysis. These matrices lack any individual-level information and, by definition, cannot be disaggregated.

This conservative approach has two major additional advantages over and above the protection of privacy and the security of the data. First, it facilitates a flexible approach to CDR analytics in the face of diverse social and public health policy problems. By working closely with researchers to determine the appropriate levels of aggregation, derived data sets can be repurposed for many different problems that might require different methodological approaches. This emphasizes the importance of setting a global standard for the process of data sharing itself, rather than a standard method of aggregating CDR data. Second, this framework serves the important purpose of bringing together cross-sector teams with different skillsets within countries that need expertise and resources. This allows for a more effective collaboration, since the operator can help with interpretation of the CDRs and the country-specific nuances of the data, and it provides an opportunity to develop in-country networks of public health practitioners, researchers, and mobile operators.

Here, we envisage the possibility of moving towards a regular transfer of data from mobile operators to trained individuals within public health and development agencies, where results can be rapidly integrated into policy measures. Indeed, the utility and sustainability of public health programs integrating these approaches will rely on successful training of local researchers in CDR analytics, and on personal relationships and particular problems rather than a one-size-fits-all model.

As human populations grow and become more mobile, these data will become an increasingly valuable tool for controlling infectious diseases, responding to natural disasters and conflict, and enabling economic growth in the poorest communities. Currently, the lack of robust protocols for protecting subscriber privacy and ensuring data security are hindering efforts to develop these approaches, however. We believe this model of analysis could serve as a useful and conservative approach for the standardization of this process, providing robust protection of individual privacy while leveraging the enormous potential of CDRs to improve human lives.

# References

1   Salathe, M. *et al.* Digital epidemiology. *PLoS Comput Biol* **8**, e1002616, doi:10.1371/journal.pcbi.1002616 PCOMPBIOL-D-12-00494 [pii] (2012).

2   Bengtsson, L., Lu, X., Thorson, A., Garfield, R. & von Schreb, J. Improved response to disasters and outbreaks by tracking population movements with mobile phone network data: a post-earthquake geospatial study in Haiti. *PLoS medicine* **8**, e1001083, doi:10.1371/journal.pmed.1001083 (2011).

3   Gething, P. W. & Tatem, A. J. Can mobile phone data improve emergency response to natural disasters? *PLoS Med* **8**, e1001085, doi:10.1371/journal.pmed.1001085 PMEDICINE-D-11-01635 [pii] (2011).

4   Palmer, J. R. *et al.* New approaches to human mobility: using mobile phones for demographic research. *Demography* **50**, 1105-1128, doi:10.1007/s13524-012-0175-z (2013).

5   Oliver, N., Matic, A. & Frias-Martinez, E. Mobile Network Data for Public Health: Opportunities and Challenges. *Front Public Health* **3**, 189, doi:10.3389/fpubh.2015.00189 (2015).

6   Tatem, A. J. *et al.* The use of mobile phone data for the estimation of the travel patterns and imported Plasmodium falciparum rates among Zanzibar residents. *Malar J* **8**, 287, doi:10.1186/1475-2875-8-287 (2009).

7   Toole, J. L., Herrera-Yaque, C., Schneider, C. M. & Gonzalez, M. C. Coupling human mobility and social ties. *J R Soc Interface* **12**, doi:10.1098/rsif.2014.1128 (2015).

8   Song, C., Qu, Z., Blumm, N. & Barabasi, A. L. Limits of predictability in human mobility. *Science* **327**, 1018-1021, doi:10.1126/science.1177170 (2010).

9   Simini, F., Maritan, A. & Neda, Z. Human mobility in a continuum approach. *PLoS One* **8**, e60069, doi:10.1371/journal.pone.0060069 (2013).

10  Lu, X., Wetter, E., Bharti, N., Tatem, A. J. & Bengtsson, L. Approaching the limit of predictability in human mobility. *Sci Rep* **3**, 2923, doi:10.1038/srep02923 (2013).

11  Wesolowski, A. *et al.* Impact of human mobility on the emergence of dengue epidemics in Pakistan. *Proc Natl Acad Sci U S A* **112**, 11887-11892, doi:10.1073/pnas.1504964112 (2015).

12  Wesolowski, A. *et al.* Quantifying seasonal population fluxes driving rubella transmission dynamics using mobile phone data. *Proc Natl Acad Sci U S A* **112**, 11114-11119, doi:10.1073/pnas.1423542112 (2015).

13  Wesolowski, A. *et al.* Quantifying the impact of accessibility on preventive healthcare in sub-Saharan Africa using mobile phone data. *Epidemiology* **26**, 223-228, doi:10.1097/EDE.0000000000000239 (2015).

14      Wesolowski, A. *et al.* Quantifying the impact of human mobility on
        malaria. *Science* **338**, 267-270, doi:10.1126/science.1223467 (2012).

15      Bengtsson, L. *et al.* Using mobile phone data to predict the spatial
        spread of cholera. *Sci Rep* **5**, 8923, doi:10.1038/srep08923 (2015).

16      Wesolowski, A. *et al.* Commentary: containing the ebola outbreak - the
        potential and challenge of mobile network data. *PLoS Curr* **6**,
        doi:10.1371/currents.outbreaks.0177e7fcf52217b8b634376e2f3efc5e
        (2014).

17      Bengtsson, L., Lu, X., Thorson, A., Garfield, R. & von Schreeb, J. Improved
        response to disasters and outbreaks by tracking population movements
        with mobile phone network data: a post-earthquake geospatial study in
        Haiti. *PLoS Med* **8**, e1001083, doi:10.1371/journal.pmed.1001083
        (2011).

18      Stoddard, S. T. *et al.* House-to-house human movement drives dengue
        virus transmission. *Proc Natl Acad Sci U S A* **110**, 994-999,
        doi:10.1073/pnas.1213349110 (2013).